\documentclass[aps,showkeys,twocolumn,superscriptaddress,prl,floatfix]{revtex4}

\usepackage{amssymb,amsmath,graphicx}
\usepackage{bm}

\begin{document}

\bibliographystyle{apsrev}
\preprint{}

\title{Tunneling Spectra of Individual Magnetic Endofullerene Molecules}

\author{Jacob E. Grose}
\affiliation{Laboratory of Atomic and Solid State Physics, Cornell University, Ithaca,
New York 14853, USA}

\author{Eugenia S. Tam}
\affiliation{Laboratory of Atomic and Solid State Physics, Cornell University, Ithaca,
New York 14853, USA}

\author{Carsten Timm}
\affiliation{Department of Physics and Astronomy, University of Kansas, Lawrence, Kansas
66045, USA}

\author{Michael Scheloske}
\affiliation{Institut f\"ur Experimentalphysik, Freie Universit\"at Berlin, Arnimallee
14, 14195 Berlin, Germany}

\author{Burak Ulgut}
\affiliation{Department of Chemistry and Chemical Biology, Cornell University, Ithaca, New York 14853,
USA}

\author{Joshua J. Parks}
\affiliation{Laboratory of Atomic and Solid State Physics, Cornell University, Ithaca,
New York 14853, USA}

\author{H\'ector~D.~Abru\~na}
\affiliation{Department of Chemistry and Chemical Biology, Cornell University, Ithaca, New York 14853,
USA}

\author{Wolfgang Harneit}
\affiliation{Institut f\"ur Experimentalphysik, Freie Universit\"at Berlin, Arnimallee
14, 14195 Berlin, Germany}

\author{Daniel C. Ralph}
\affiliation{Laboratory of Atomic and Solid State Physics, Cornell University, Ithaca,
New York 14853, USA}

\date{\today}


\begin{abstract}
The manipulation of single magnetic molecules may enable new strategies for high-density
information storage and quantum-state control.  However, progress in these areas
depends on developing techniques for addressing individual molecules and controlling
their spin. Here we report success in making electrical contact to individual magnetic
N@C$_{60}$ molecules and measuring spin excitations in their electron tunneling spectra.
We verify that the molecules remain magnetic by observing a transition as a function of
magnetic field which changes the spin quantum number and also the existence of
nonequilibrium tunneling originating from low-energy excited states.  From the tunneling
spectra, we identify the charge and spin states of the molecule.  The measured spectra
can be reproduced theoretically by accounting for the exchange interaction between the
nitrogen spin and electron(s) on the C$_{60}$ cage.
\end{abstract}

\maketitle

Magnetic molecules provide the opportunity to study, at a fundamental level, the
origins of ferromagnetism and quantum aspects of magnetic dynamics \cite{1,2}. Proposals
suggest that their spin degrees of freedom could also serve as useful qubits for
quantum manipulation \cite{3}. Most previous experiments on magnetic molecules have
examined samples consisting of many molecules; however, applications and more
detailed scientific studies will require the ability to address individual
molecules. Here we report the use of tunneling spectroscopy within single-molecule
transistors (SMTs), a technique which has been used previously to measure vibrational
and electronic excitations \cite{4,5,6}, to achieve electrical contact to individual
molecules of the spin-3/2 endohedral fullerene N@C$_{60}$ (Fig. 1a) and to measure its
spin excitations. N@C$_{60}$ is an attractive model system \cite{7,8} because of its
simple spin structure and because of the possibility of control	experiments employing
nonmagnetic C$_{60}$ molecules \cite{4,9,10,11,12,13,14}.  N@C$_{60}$ molecules also
have the advantage of being stable \cite{15} at the high temperatures present during
the electromigration process by which our molecular-scale junctions are formed \cite{16,
17}.  Previous SMT experiments using less robust Mn$_{12}$-based magnetic molecules
detected magnetic signatures in a few samples \cite{18,19}, but	found that the molecular
magnetism was usually destroyed during device fabrication \cite{19}. We observe that the
N@C$_{60}$ devices exhibit clear magnetic character, in that they have a spin-state
transition as a function of applied magnetic field.  The nature of this transition
allows us to identify the charge and spin states of the molecule in the SMT.  The
spectra of N@C$_{60}$ also exhibit low energy excited states and signatures of
nonequilibrium spin excitations	predicted for this molecule \cite{20}. We associate the
existence of a spin transition in N@C$_{60}$ accessible at laboratory magnetic fields
with the scale of the exchange interaction between the nitrogen spin and electron(s) on
the C$_{60}$ cage.

\begin{figure}[b]
	\label{Figure1}
\includegraphics[width=8.6cm]{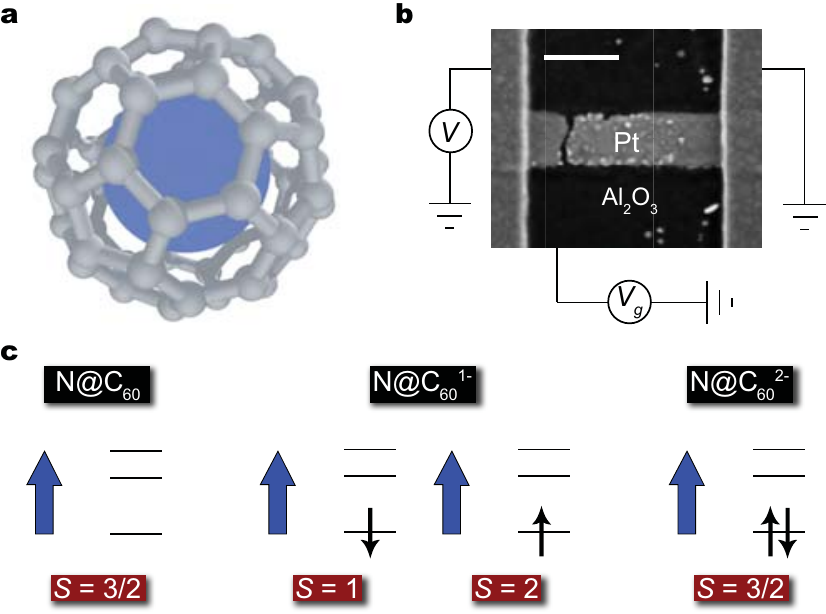} \caption{ \footnotesize{Device geometry, and the
spin states of N@C$_{60}$.  {\bf a}, Schematic diagram of the N@C$_{60}$ molecule. {\bf
b}, Scanning electron microscopy image of a sample at room temperature following
electromigration. The scale bar denotes 200~nm.  The circuit schematic shows
our biasing convention.  {\bf c}, Spin states for N@C$_{60}$ and its anions. The blue
arrows represent the $S_N$~=~3/2 spin of the N atom.  The horizontal black lines
represent the lowest unoccupied molecular orbitals of the C$_{60}$.  The possible total
spin states of the N@C$_{60}$ ion are indicated.  }}
\end{figure}

The geometry of our SMT devices is shown in Fig.\ 1b.  We make the devices by
adsorbing the molecules onto an initially-continuous Pt wire on top of an oxidized Al
gate electrode and then breaking the wire using electromigration to form a nm-scale gap.
(Additional information about fabrication procedures is	given in the Methods section.)
We identify the presence of one or more molecules in the gap by the observation of
gate-dependent Coulomb-blockade transport characteristics \cite{5,11}.  Our success rate
for obtaining such devices was 9/19 for N@C$_{60}$ devices and 17/59 for C$_{60}$
devices, while 0/39 control samples prepared using pure toluene instead of a fullerene solution
showed Coulomb blockade.  Six of the N@C$_{60}$ devices were sufficiently stable for
detailed measurements.

\begin{figure} [t]
	\label{Figure2}
\includegraphics[width=8.6cm]{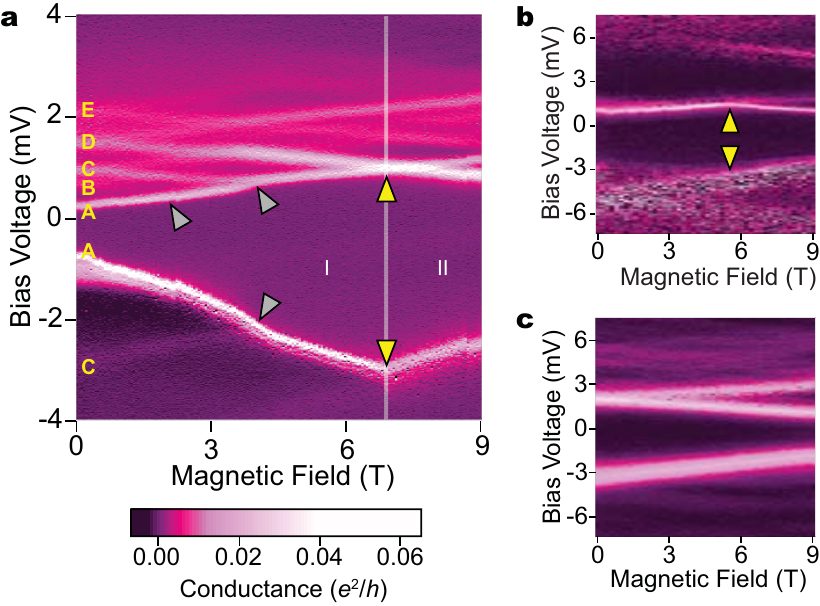} \caption{ \footnotesize{Color-scale plots of
differential conductance ($dI/dV$) for N@C$_{60}$ and C$_{60}$ single-molecule
transistors.  {\bf a}, $dI/dV$ for N@C$_{60}$ device \#1 as a function of source-drain
voltage and applied magnetic field at constant value of gate voltage
$V_g$~=~-891 mV on the positive side of the degeneracy point. Yellow triangles indicate
a change in slope of the ground-state peaks, and gray triangles mark the non-crossing
levels discussed in the text. The labels designate different transitions.  In region I,
the ground-state transition corresponds to decreasing spin, $S_{z,q+1}$ $-$ $S_{z,q}$ $<$ 0,
while in region II $S_{z,q+1}$ $-$ $S_{z,q}$ $>$ 0.  {\bf b}, Same as {\bf a} for N@C$_{60}$
device \#2.  The maximum of the color scale is 0.15 $e^{2}/h$.  {\bf c}, $dI/dV$ as a
function of $V$ and $B$ at constant $V_g$ on the positive side of the degeneracy point
for a C$_{60}$ single-molecule transistor. The maximum of the color scale is 0.02
$e^{2}/h$.  No change in slope is observed for the ground-state peaks of C$_{60}$.}}
\end{figure}

The most striking difference between the N@C$_{60}$ devices and the C$_{60}$ controls is
seen in the magnetic field dependence of the tunneling conductance spectra.  In Fig.~2
we show these spectra as a function of source-drain voltage ($V$) and magnetic field
($B$) at dilution-refrigerator temperatures (electron temperature $\approx$ 100 mK) for
values of gate voltage ($V_g$) slightly more positive than the ``degeneracy point'' --
the value of gate voltage for which the energy difference between the two accessible
charge states is zero, so that current can flow even near $V$~=~0.  Figure 3 shows the
conductance for each sample as a function of $V$ and $V_g$ near the degeneracy point.
(By focusing on low-voltage transport near one degeneracy point, we ensure that
tunneling is occurring through a single molecule \cite{5}.)  In the N@C$_{60}$ devices,
as a function of increasing $B$ (Fig.\ 2a,b) the lowest-voltage conductance peaks at
positive and negative bias first move apart and then change slope to move back closer
together. These initial peaks can be associated with the same quantum transition from
the ground state of one charge state (charge $q$) of the molecule to the ground state
with charge $q+1$ (one fewer electron).  At the value of $B$ where the slopes of the
peaks change
sign in Fig.\ 2a, we observe level crossings, where a conductance peak that at
low $B$ is associated with an excited state of charge $q+1$ becomes the ground state
of charge $q+1$ at high $B$. We measured ground-state conductance peaks that shift
apart then together as a function of $B$ on the positive-gate side of the
degeneracy point in four out of five of the N@C$_{60}$ devices on which we performed
this measurement.  (The fifth showed no change in slope up to 9 T.)  The field values at
which the crossovers occurred varied from approximately 1 T to 7 T.

The level crossing in Fig.\ 2a shows that the change in sign of the
ground-state-peak slopes is due to an increase in total spin component  along the
magnetic field direction  for the lowest-energy charge $q+1$ state. The slopes of the
peaks as a function of $B$ imply that at low fields $S_{z,q} > S_{z,q+1}$, and at high
fields $S_{z,q} < S_{z,q+1}$.
Assuming that $\left|S_{z,q+1} - S_{z,q}\right| = 1/2$ , the slopes of the
conductance resonances vs. magnetic field shown in Fig.\ 2 correspond to an electronic
$g$-factor $|g| = 2.0 \pm 0.3$.

In contrast, Fig.\ 2c shows the tunneling spectrum for a C$_{60}$ device also taken on
the positive-gate side of its degeneracy point. Of the five C$_{60}$ devices we
measured as a function of magnetic field, none showed a change in the sign of the
slopes for the lowest-voltage tunneling transition.  While the spectra for the C$_{60}$
devices often contain low-energy excitations in the same range of energy as those of
N@C$_{60}$, 0.1 meV to several meV, the lack of spin transitions suggests that the
low-lying excitations in C$_{60}$ are vibrational satellites \cite{4,11,21}, rather
than being electronic or magnetic states.

Recently, a tunneling transition with a change in slope as a function of magnetic
field has been observed in a C$_{60}$ device, for gate voltages in a limited range
near the middle of one Coulomb-blockade region \cite{14}. Our measurements showing no
such level crossing in C$_{60}$ devices are not in conflict with ref. [14] because our
data for C$_{60}$ and N@C$_{60}$ likely correspond to a different charge state, and our
measurements were made only for gate voltages near the degeneracy point, well
away from the middle of any Coulomb-blockade region.

We assert that the spin-state transition in the N@C$_{60}$ devices is likely
associated with two states with different values of total spin, as opposed to
two states within the same multiplet split by magnetic anisotropy.  Because any
magnetic anisotropy axes could be oriented arbitrarily relative to the applied
magnetic field, if the transition were due to states split by magnetic
anisotropy one should expect, in general, that the slopes of the energy levels
as a function of $B$ should show deviations from $|g| = 2$ and exhibit strong
deviations from linearity at low field \cite{19,22,23}.  In our data, the field
dependence of the transitions is very close to linear (away from localized kinks
associated with level crossings).  The lack of observable magnetic anisotropy
effects is consistent with the symmetries of N@C$_{60}$ and the weak spin-orbit
coupling of its constituent atoms \cite{24}.

\begin{figure}[t]
       \label{Figure3}
       \includegraphics[width=8.6cm]{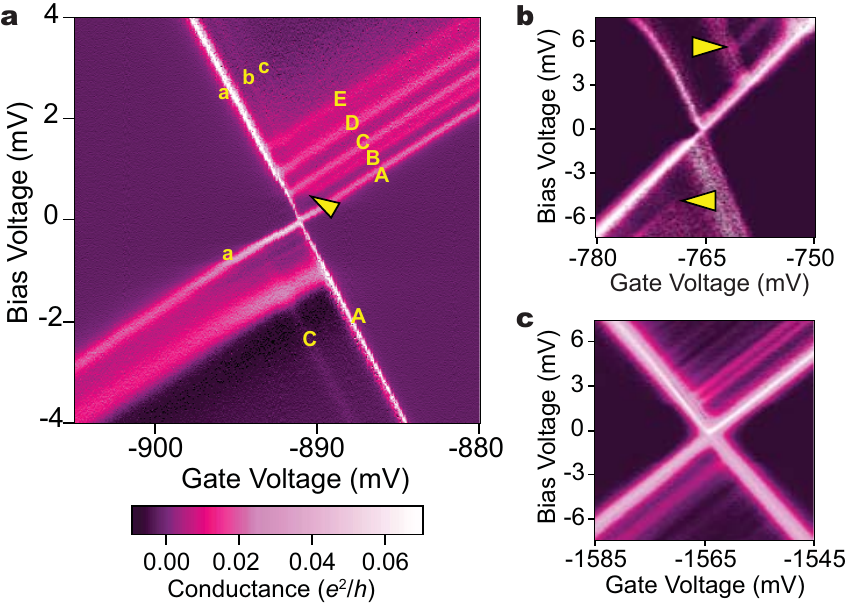}
\caption{ \footnotesize{Color-scale plots of differential conductance ($dI/dV$) as a
function of bias voltage and gate voltage at zero applied field ($B$ = 0 T).  {\bf a,b},
$dI/dV$
from the same N@C$_{60}$ devices \#1 and \#2 shown in Fig. 2a and b, respectively. The
labels in {\bf a} designate different transitions, consistent with Fig. 2a. Their
energies, determined
by extrapolating the transition lines to the Coulomb-blockade threshold, are ($\pm$ 0.02
meV) $E_B-E_A$ = 0.29 meV, $E_C-E_A$ = 0.48 meV, $E_D-E_A$ = 0.93 meV, $E_E-E_A$ = 1.35
meV, $E_b-E_a$ = 0.20 meV and $E_c-E_a$ = 0.41 meV.  The triangles indicate the
termination of ETE peaks
associated with non-equilibrium excitations.  {\bf c}, $dI/dV$ for the same C$_{60}$
device shown in Fig. 2c.  No ETE peaks are visible for this nonmagnetic molecule. }}
\end{figure}

With this understanding, the existence of a spin-state transition at an
accessible magnetic field for the charge $q+1$ energy levels of N@C$_{60}$ allows us to
identify the charge and spin states involved in the tunneling.  Consider the
possible states shown in Fig.\ 1c, which depicts the nitrogen spin (blue arrow)
coupled to electron spins (black arrows) occupying the lowest unoccupied
molecular orbital (LUMO) of the C$_{60}$.  We assume that a 3-fold degeneracy of the
LUMOs of isolated C$_{60}$ is broken by molecular deformation due to interaction with
the electrodes or the Jahn-Teller effect \cite{25}, and we consider only the
experimentally accessible neutral and negative charge states. Neutral N@C$_{60}$ is
known to have spin 3/2 because the encapsulated nitrogen retains its atomic
electronic configuration \cite{26}. When electrons are added to N@C$_{60}$, they are
believed to occupy the LUMO of the C$_{60}$ molecule \cite{27}.  With the addition of
one extra electron, N@C$_{60}^{1-}$ may therefore have either $S$ = 1 or 2, and these
two multiplets are split in energy by the exchange interaction between the N spin and the electron
spin.  If the exchange is antiferromagnetic, the ground state will have $S$ = 1
for low $B$ and $S$ = 2 for large $B$.  For N@C$_{60}^{2-}$, the total spin should again
be $S$ = 3/2 since the splitting between the LUMO states is expected to be large
compared to
the Hund's rule coupling.  Because we observe the spin-state transition in the
{\it more positive} of the two charge states involved in tunneling, the 0/1- charge
couple can be ruled out as the charge states between which tunneling occurs.
This leads to the identification, for an antiferromagnetic exchange interaction,
that 1-/2- is the couple with the lowest charge magnitude that is capable of
transitioning from $S_{q} - S_{q+1}$ = +1/2 at low fields to $S_{q} - S_{q+1}$ = $-$1/2 at
high fields. For a ferromagnetic exchange interaction, the lowest-magnitude charge
couple possibly consistent with the data would be 2-/3-, which is unlikely
because the 3- charge state should have a much higher energy than 1- or 2-
states.  A 2-/3- scenario would also require fine tuning such that the exchange
splitting is approximately equal to the deformation-induced level splitting
within the N@C$_{60}^{2-}$ LUMO.

The identification of our measured spectra with the 1-/2- charge couple implies
that our N@C$_{60}$ devices are in the 2- charge state at $V_g$ = 0, which may seem to
be a large magnitude of charge in an unbiased device.  However, for C$_{60}$ in the gas
phase, both the 1- and 2- states are more stable than the neutral form, with the
2- state quite close in energy to the 1- state \cite{28,29,30}.  The equilibrium charge
of C$_{60}$ is generally found to be either 1- or 2- when it is adsorbed on noble
metal surfaces \cite{31,32}.  These previous studies suggest that the 3- charge state is
likely to have a much higher energy than 1- or 2- in an unbiased device, which
argues against the possibility of a ferromagnetic exchange interaction in our
samples as this would require a 2-/3- charge couple with a 3- charge state at
zero gate voltage.  In regard to the sign of the exchange interaction, an
{\it ab-initio} calculation has predicted a weakly ferromagnetic interaction in
N@C$_{60}^{1-}$, rather than the antiferromagnetic interaction we suggest \cite{33}.
However, it should be
noted that the calculation of the exchange interaction is difficult since it is
computed as a small difference between large total energies.

The device-to-device variations that we measure in the N@C$_{60}$ tunneling spectra
indicate that the energies of the electronic states are perturbed differently by
the environment in each device.  In order to explain the variations in field
value giving rise to the spin-state transition, the strength of the exchange
interaction must differ between devices, presumably due to different amounts of
molecular deformation and/or local electric fields.  If we parameterize the
exchange interaction in the form $J \mathbf{s}_{e} \cdot \mathbf{S} _{N}$, where
$\mathbf{s}_{e}$ is the total electron spin and $\mathbf{S} _{N}$ is the
spin on the nitrogen atom, then measured transition fields (1 to 7 T) imply
exchange strengths, $|J|$ , between 0.06 meV and 0.4 meV.

\begin{figure*}
       \label{Figure4}
       \includegraphics[width=14.0cm]{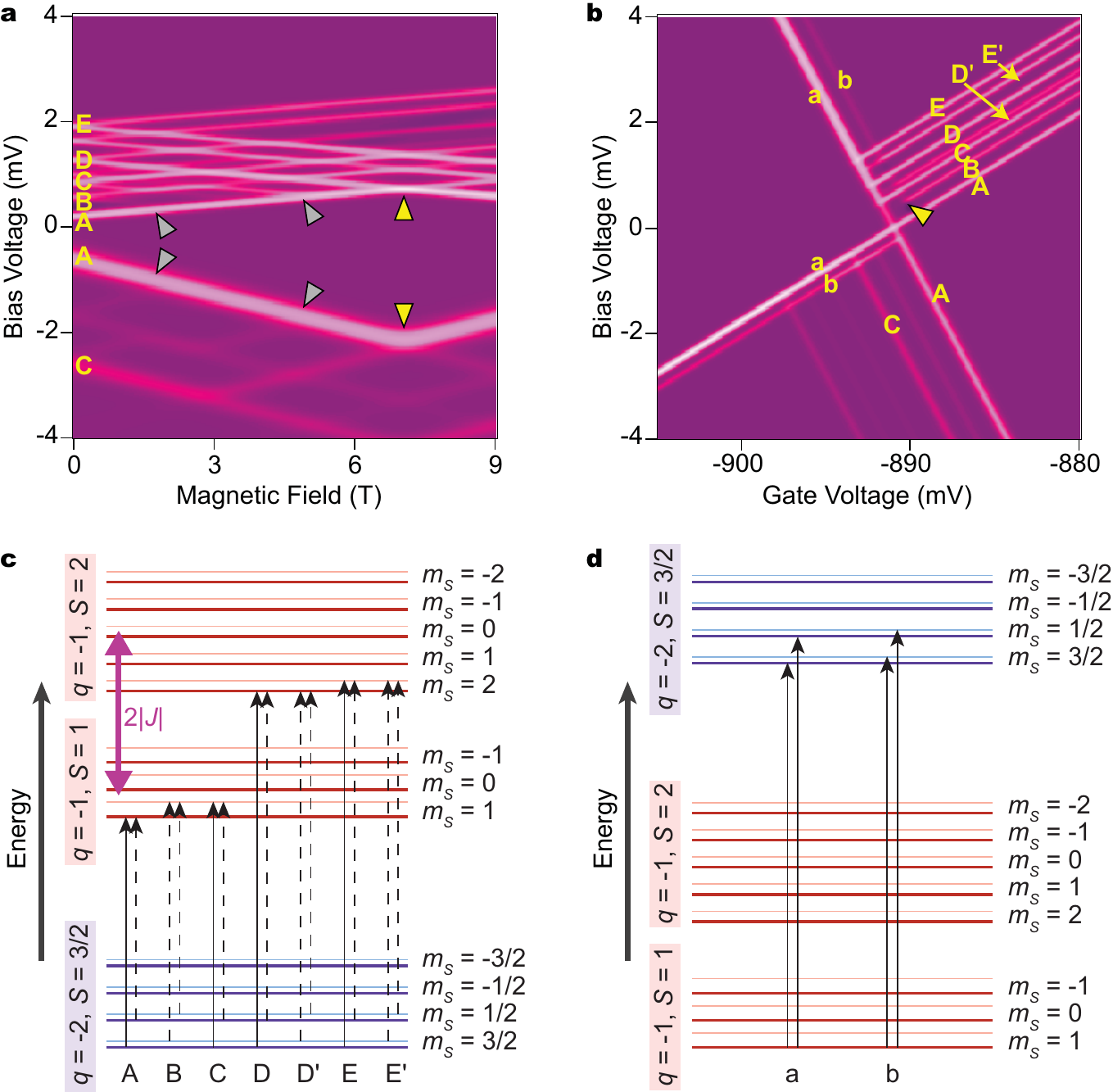}
\caption{ \footnotesize{Numerical calculation of the tunneling spectrum for N@C
$_{60}$, with parameters chosen to mimic the data for Device \#1.  {\bf a}, Calculated
$dI/dV$ as a function of $V$ and $B$ at constant $V_g$ on the positive side of the $q$ =
-2 / $q$ = -1 degeneracy point. Compare to Fig.\ 2a.  The yellow triangles mark the
spin-state transition, and the gray triangles
mark non-crossing levels.  {\bf b}, Calculated $dI/dV$ as a function of $V$ and $V_g$.
The triangle
marks the termination of a nonequilibrium ETE peak.  Compare to Fig.\ 3a.  {\bf c},
Energy
level diagram for a large applied magnetic field showing allowed tunneling transitions
which contribute to the low-energy spectrum, for tunneling from the charge $q$ = -2
states
(blue) to the $q$ = -1 states (red) of N@C$_{60}$.  The dotted lines show nonequilibrium
transitions.  (For $B$ = 0, the different $m_S$ levels within a given multiplet will
become
degenerate in the absence of magnetic anisotropy.) {\bf d}, Allowed tunneling
transitions
contributing to the low-energy spectrum, for tunneling from $q$ = -1 states to $q$ = -2
states. The letters labeling the transitions in panels c and d can explain the labeled
states in Figures 2a, 3a.  The relative energies of the $q$ = -1 and $q$ = -2 states
will
depend on $V_g$.}}
\end{figure*}

To understand how the measured tunneling spectra might relate to the molecular
magnetism, we have calculated the conductance of an N@C$_{60}$ transistor (Fig. 4a,b).
As a minimal model to account for the low-energy transitions, we consider the
energy level diagrams shown in Fig. 4c,d, which include the three multiplets,
($q$ = -2, $S$ = 3/2), ($q$ = -1, $S$ = 1), and ($q$ = -1, $S$ = 2), we have introduced
previously and also the lowest-energy excited-state multiplet associated with
each of these states.  The excited multiplets could correspond to the added
excitation of a long-lived vibrational mode or energy shifts due to charging in
nearby molecules. We assume that the molecule has negligible magnetic anisotropy,
so that all of the magnetic effects we consider are due to exchange-induced level
splittings.  This is in contrast to previous calculations for Mn$_{12}$-type molecules
where the low-energy spectra were assumed to be dominated by large magnetic
anisotropy \cite{18,19}. We use the rate-equation approach \cite{20} to calculate the
differential conductance.  Parameter values, listed in the Methods section, are
selected to approximate the data in Fig.\ 2a and 3a.

There are a sufficiently large number of free parameters that we do not seek or
claim quantitative agreement.  Depending on assumptions about the value of $J$ and
vibrational parameters, different numbers of excited-state tunneling transitions
are possible.  Nevertheless the calculation can reproduce many of the
qualitative features seen in Fig.\ 2 and 3.  First, the N@C$_{60}$ spectra shown in
Fig.\ 3a,b and the theoretical results shown in Fig.\ 4b both exhibit a pattern of
nonequilibrium excitations that was predicted \cite{20} for spin excitations in
N@C$_{60}$.  If one extrapolates the conductance peaks marked by yellow triangles in
Fig. 3a and b toward zero voltage, they do not extend all the way to the line
corresponding to the ground-state tunneling threshold, but instead terminate
when they intersect a conductance peak corresponding to an excited-state
transition.  We observed this type of feature in four of six N@C$_{60}$ devices, but
in none of thirteen C$_{60}$ devices that were stable enough for detailed study
({\it e.g.}, Fig.\ 3c). This type of phenomenon, which has been measured previously in
metal and semiconductor quantum dots \cite{34,35}, can be understood \cite{20} in terms
of excited-state-to-excited-state (ETE) transitions within the spectrum of energy
levels in N@C$_{60}$.  The idea is illustrated by Fig.\ 4c,d.  Here ground-state
tunneling at low magnetic field corresponds to the transition (labeled A) from
the $q$ = -2, $S$ = 3/2 state to $q$ = -1, $S$ = 1. For a sufficiently large applied
voltage $V$, the tunneling of an electron back into the molecule can lead to a
transition (labeled b) from the $q$ = -1, $S$ = 1 state to the excited $q$ = -2,
$S$ = 3/2
multiplet, rather than the ground state $S$ = 3/2 multiplet.  If this excited
state does not relax before the next electron tunnels out of the molecule, this
will produce an ETE conductance peak corresponding to a transition (B) between
the excited $q$ = -2, $S$ = 3/2 multiplet and the excited $q$ = -1, $S$ = 2 multiplet.
The reason that transition B cannot extrapolate indefinitely to lower voltage is
that the applied bias must provide sufficient energy to permit the transition b
for transition B to be possible.  As support for this interpretation, note that
the level diagram in Fig. 4c,d requires that $E_C = E_B + (E_b - E_a)$.   Using the
measured values of the excitation energies (see the caption for Fig.\ 3), this
equality is satisfied within experimental uncertainty.

Another successful feature of the calculation is that it can account for the
magnetic-field dependence of the tunneling spectrum shown in Fig.\ 2a.  In
particular, note that this spectrum has two conductance peaks that as a function
of magnetic field intersect the lowest-voltage transition but then terminate so
that they do not produce a level crossing -- this happens near 1.5 T for
transition B and 4 T for transition C.  Only near 7 T is there a true level
crossing for transition D that corresponds to the spin-state transition discussed
above.  The ``non-crossing'' levels can be understood as another consequence of
nonequilibrium ETE transitions. For voltages less than the ground-state to
ground-state transition, the first electron tunneling event that initiates
current flow is not allowed, so that the excited states cannot be populated.

One difference between the calculated spectrum in Fig.\ 4a,b and the measurements
in Fig.\ 2a and 3a is that the calculated spectrum contains two more lines at
positive bias (7 lines at $B$ = 0 instead of 5).  The extra lines (D$'$ and E$'$)
correspond to non-equilibrium ETE transitions from the excited $q$ = -2, $S$ = 3/2
state to the two $q$ = -1, $S$ = 2 states (i.e., they are lower-energy satellites
of transitions D and E).  We suspect that these extra transitions are present but
unresolved in the experimental spectra because the linewidths of the experimental
transitions increase as a function of energy and the ETE transitions have weaker
amplitudes than transitions originating from the ground state, thereby making
them difficult to distinguish at higher energies.  The level broadening with
increasing energy is not included in our simple model, but if we add in the
measured broadening by hand, the transitions D$'$ is completely obscured and
E$'$ becomes only marginally resolvable.

The capability to make electrical contact to single magnetic molecules without
destroying their magnetic character provides the foundation for more advanced
strategies to manipulate spin states.  We have shown that the charge and spin
values of the molecule can be identified, that spin excitation energies can be
measured, and that the tunneling spectrum of N@C$_{60}$ can be understood
qualitatively within an elementary model.  The next challenges will include
developing the means to coherently populate desired superpositions of spin
states (using, for instance, applied microwaves \cite{3,36}), developing
accurate techniques to read out spin information, and achieving improved control over the
interaction between the magnetic molecule and its environment so as to minimize
device-to-device variations.

\section{METHODS}
\noindent{\bf Synthesis}: N@C$_{60}$ was synthesized by continuous nitrogen ion
implantation into
freshly
sublimed fullerene layers \cite{26} with a yield (N@C$_{60}$:C$_{60}$ ratio) of
$\sim$10$^{-4}$. The N@C$_{60}$ contained
in the harvested product was enriched and purified by multi-step high-pressure liquid
chromatography (HPLC) \cite{37}. Purity was checked by UV-Vis absorption, HPLC, and
electron spin resonance and found to be better than 99.5\%.\\

\noindent{\bf Electrochemistry}: We have confirmed, using electrochemical measurements,
that the
interaction in N@C$_{60}$ between the N atom and the extra electrons on C$_{60}$ is
relatively
weak.  Cyclic voltammograms were performed on N@C$_{60}$ or C$_{60}$ films formed by
drop casting
from toluene solutions onto a 3 mm Pt disc electrode, polished to a mirror
finish, used as the working electrode.  A Ag/AgCl (saturated NaCl) and a Pt coil were
used as the reference and counter electrodes, respectively.  The scans, carried out at a
sweep rate of 500 mV/s, were done in 0.1 M tetrabutylammonium hexafluorophosphate in
acetonitrile in a single-compartment cell.  The solution was deoxygenated via N$_2$
bubbling for 10 min before the sweeps, which were run in a N$_2$ atmosphere.  We found
that the first two reductions of N@C$_{60}$ occur within 25 mV of those of C$_{60}$.  The ~25 mV
shifts are not related to the nature of the molecules and can be attributed to the varying
rates of dissolution of the reduced species due to slight variations in the film
morphology.  This arises as a result of the fact that while the neutral forms of
C$_{60}$ and N@C$_{60}$ are insoluble in acetonitrile, the reduced forms are not.\\

\noindent{\bf Sample Fabrication}: Our single-molecule transistors were fabricated
following aprocedure similar to those used previously \cite{4,5}. First we made an Al gate
electrode 16 nm thick and 2 $\mu$m wide on an oxidized Si wafer, and exposed the Al to
air to form a thin insulating oxide. On top of the gate electrode, we fabricated continuous Pt wires
with widths of approximately 150 nm and thicknesses of 10 nm. The chips were cleaned with an
oxygen plasma and immediately covered with 25 $\mu$L of either a 0.1 mM solution of
N@C$_{60}$ for 2.5 minutes or (for one type of control sample) a 0.5 mM solution of
C$_{60}$ in toluene for 1 minute. Then the excess solution was blown off the chip, and the deposition
process was repeated. We found that repeating the deposition process produced a more
convenient yield of single-molecule devices. After the molecules were deposited, we
cooled to cryogenic temperatures and broke the wires using electromigration in a circuit
with small series resistance \cite{17}, thereby forming nm-scale gaps in which a
molecule was sometimes trapped.  The breaking voltages for our Pt wires were in the
range 1-1.5 V.\\

\noindent{\bf Low-temperature measurements}:  The measurements were performed in a
dilution refrigerator equipped with room-temperature low-pass filters and low-temperature copper
powder filters on all electrical lines.  The electron temperature calibrated on quantum
dot samples is $< 100$ mK.  The DC current was measured as a function of swept source-
drain voltage at fixed gate voltage and magnetic field (without a lock-in amplifier),
and then the value of the gate voltage or magnetic field was stepped.  The differential
conductance was calculated by numerical differentiation.\\

\noindent{\bf Theory}: The differential conductance $dI/dV$ is calculated using the
rate-equation approach in the sequential-tunneling approximation \cite{20}.  The electronic molecular
Hamiltonian reads
\[
\mathcal{H}_{el} = (\varepsilon - eV_g^{*}) \sum _{\sigma}
a^{\dagger}_{\sigma}a_{\sigma}
+
Ua^{\dagger}_{\uparrow}a_{\uparrow}a^{\dagger}_{\downarrow}a_{\downarrow} - J\mathbf{s}_
{e} \cdot \mathbf{S}_{N} - B(s_{e}^{z} + S_{N}^{z})
\]
where $\varepsilon$ is the on-site energy of electrons with spin
$\sigma = \uparrow,\downarrow$ created by $a_{\sigma}^{\dagger}$, $eV_g^{*}$ is the
electric
potential at the position of the molecule, $U$ is the Coulomb repulsion, $J$  is the
exchange
interaction between the nitrogen spin $\mathbf{S}_{N}$ and the spin
$\mathbf{s}_{e} = \sum_{\sigma\sigma'} a^{\dagger}_{\sigma}
(\mbox{\boldmath$\sigma$}_{\sigma,\sigma'}/2)a_{\sigma'}$ of
the electrons in the LUMO, and $B$
is the magnetic field with a factor $g\mu_{B}$ absorbed. Only one LUMO is included, the
other two
are assumed to be split off due to molecular deformation and to not to affect $dI/dV$.
The asymmetric contact capacitances and the fact that the bias voltage is applied to the
(arbitrarily labeled) left electrode are incorporated by writing
$V_{g}^{*} = \alpha V_{g} + \beta_{L}V$  with the applied gate
voltage $V_{g}$ and bias voltage $V$, where $\alpha$ = 0.15 and $\beta_{L}$ = 0.25 give
the best fit.
Asymmetric tunneling amplitudes $t_{L,R}$ between the LUMO and the left and right
electrodes are assumed; a ratio $t_{L}/t_{R}$ = 0.4 is used. In addition, we include
low-energy excitations of energy $\hbar\omega_{n}$ in the $n$- charge state,
which in this description for the sake of definiteness we assume to be vibrational. The
transition rates between vibrational states contain Franck-Condon matrix elements
\cite{38} $M_{nm, n'm'}$ (with electron numbers $n,n'$ and oscillator quantum numbers
$m,m'$) with
relative values
of the
nonvanishing components $M_{20,10} = M_{20,11} = M_{21,11} = M_{22,10} = 1$ and
$M_{21,10}$ = 0.7; the matrix is
symmetric. The other parameters used in the
calculations are $\varepsilon$ = -3.23415, $U$ = 3.1, $J$ = -0.0004, and $k_{B}T$ =
0.00002 all in eV.

\section {Acknowledgements}

We thank R.\ D\"oring and O.\ B\"a{\ss}ler for their work to synthesize and purify the
N@C$_{60}$ and G.\ R. Hutchison for help with calculations.  The research at Cornell was
supported by the US NSF (DMR-0520404, DMR-0605742, EEC-0646547, CHE-0403806 and through
use of the Cornell Nanofabrication Facility/NNIN). Work in Berlin was supported by the
{\it Bundesministerium f\"ur Bildung und Forschung} under contract no. 03N8709.

\section {Author Contributions}
JEG played the primary role in fabricating the samples, performing the measurements, and
analyzing the data, with assistance from EST and JJP and advice from DCR.  CT performed
the model calculations.  MS and WH led the molecular synthesis, purification, and
characterization.  BU and HDA performed electrochemical characterization.  All of the
authors contributed to the data analysis and the preparation of the manuscript.

\section {Author Information}
Correspondence should be addressed to DCR.

\end{document}